\documentclass[12pt]{article}

\usepackage{graphicx}

\begin{document}

\title{\bf Long-range dependencies \\ in heart rate signals- revisited}

\author{Danuta Makowiec$^{1,3}$,  Rafa{\l } Ga{\l }{\c a}ska$^2$,\\
 Aleksandra Dudkowska$^1$, Andrzej Rynkiewicz$^2$,  Marcin Zwierz$^1$,\\
$^1${\small Institute of Theoretical Physics and Astrophysics,Gda\'nsk University, Poland} \\
$^2${\small 1st Department of Cardiology, Medical University of Gda\'nsk,  Poland }\\
$^3${\small e-mail: {\it fizdm@univ.gda.pl}}}
\date{\today}
\maketitle

\noindent{ PACS numbers: 87.19.Hh, 05.40.-a, 87.80.Vt, 89.75.Da}

\vspace{0.2in}

\centerline{\bf ABSTRACT}

{\noindent The RR series extracted from human electrocardiogram signal (ECG) is considered as a fractal stochastic process. The manifestation of long-range dependencies is the presence of power laws in scale dependent process characteristics. Exponents of these laws: $\beta$ - describing power spectrum decay, $\alpha$ - responsible for decay of detrended fluctuations or $H$ related to, so-called, roughness of a signal, are known to  differentiate hearts of healthy people from hearts with congestive heart failure. There is a strong expectation that resolution spectrum of exponents, so-called, local exponents in place of global exponents allows to study  differences between hearts in details. The arguments are given that local  exponents obtained in multifractal analysis by the two methods: wavelet transform modulus maxima (WTMM) and multifractal detrended fluctuation analysis (MDFA), allow to recognize the following four stages of the heart: healthy and young, healthy and advance in years,  subjects with left ventricle systolic dysfunction (NYHA I--III class) and characterized by severe congestive heart failure (NYHA III-IV class).}

\section{Introduction}
\begin{figure}
\includegraphics[width=0.35\textwidth]{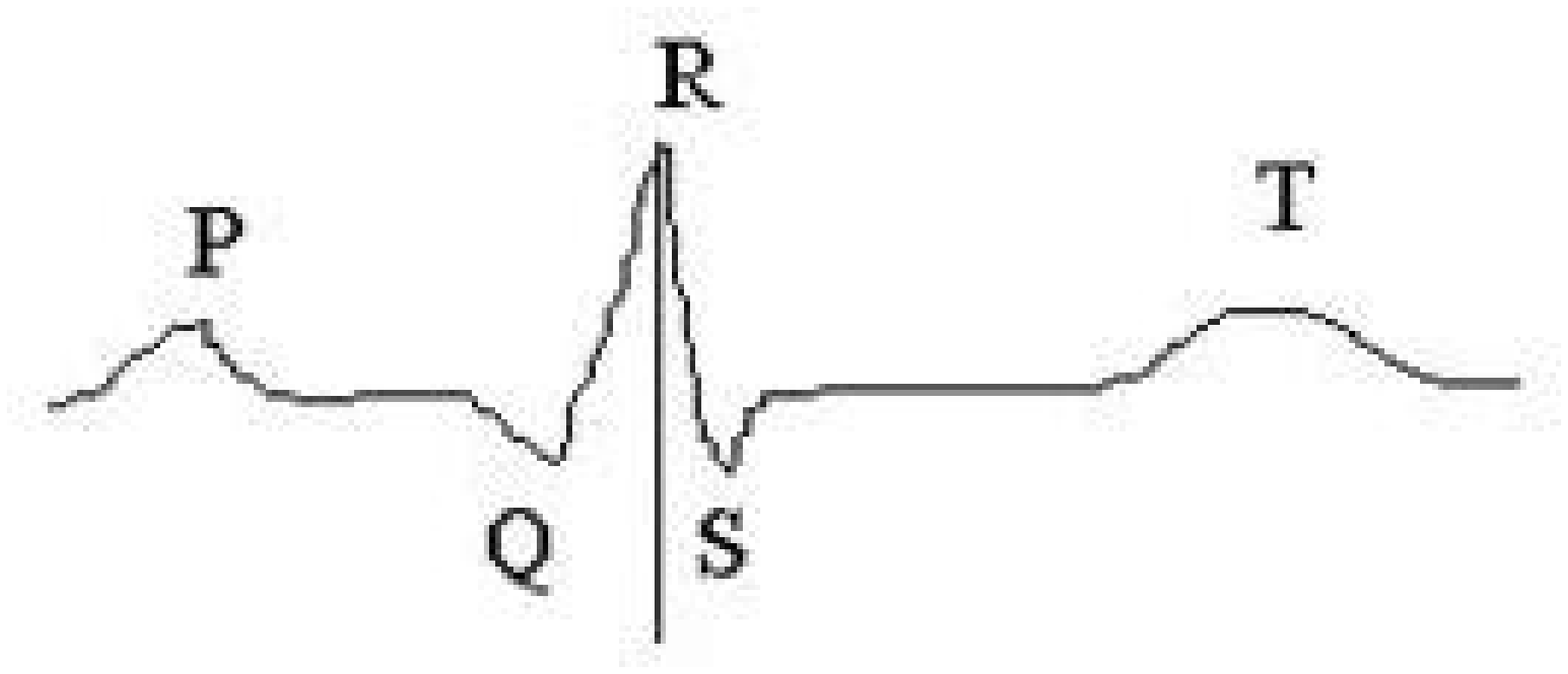}
\includegraphics[width=0.59\textwidth]{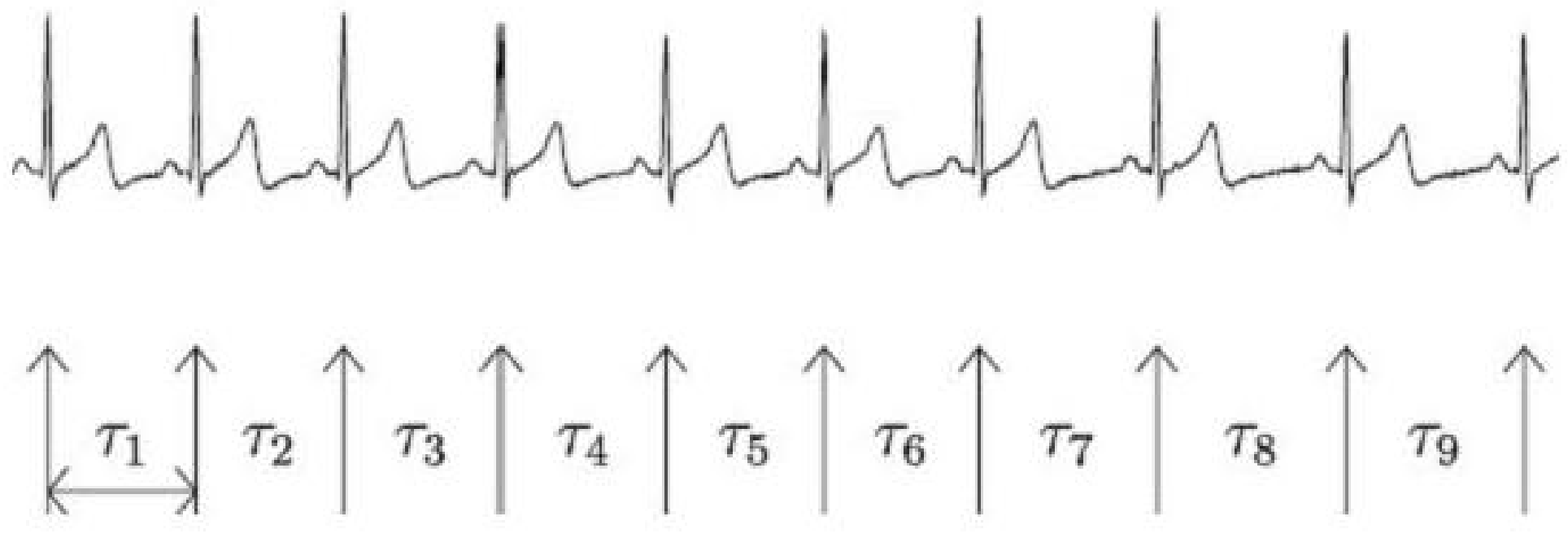}
\includegraphics[width=0.99\textwidth ]{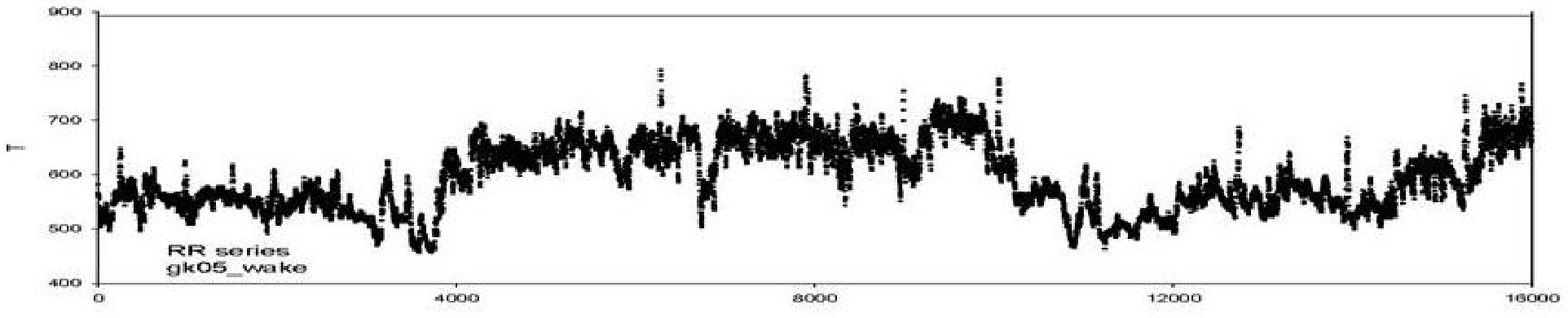}
\caption{\label{fig1}Electrocardiogram analysis: (a)a single P-QRS-T cycle (b) identification of  RR-intervals (c)a heartbeat time series from a healthy individual in a daily activity: a normal RR signal.  }
\end{figure}

Heart rate variability (HRV) represents the most promising marker for measuring activity of the autonomic nervous system --- the  system that is responsible for cardiovascular regulation in normal and pathologic conditions and is related to mortality \cite{guide}.  The wide popularity of HRV study is ensured  by the noninvasive, easily obtainable techniques providing a signal to analysis \cite{guide,Goldberger,Glass,Teich,SpectralScaling}. The, so-called, electrocardiogram (ECG), a recording of cardiac-induced potential,  reveals the basic information about atrial and ventricular electrical activity of the heart, see Fig \ref{fig1}. Readily recognizable features of ECG are labeled by the letters $P$ - $QRS$ - $T$. In a continuous ECG  record  each QRS complex   is detected, and  the so-called normal-to-normal (NN) intervals between subsequents $R$ peaks are determined. 

In mathematical terms, the heartbeat signal is modeled as a point process. The occurrence of a contraction at time $t_i$ is represented by an impulse of Dirac delta function $\delta(t-t_i)$, so that the sequence of heartbeats is as follows
\begin{equation}
 X(t) =\sum_i \delta(t-t_i)
 \end{equation}
Any realization of a point process is specified by the set of occurrence times $\{t_i\}$ of these events. The RR-signal $\{\tau_i\}$ is a sequence of positive numbers representing adjacent times $\tau_i=t_{i+1}-t_i$. 
(The ectopic beats or arrhythmic events which are present in the ECG record and which are not represented in RR series are beside the scope of our investigations, though we remember about their role in quantifying the heart state, see, e.g., \cite{arrhythmia}.) 

Clinicians say about normal activity of the heart as regular sinus rhythm. But normal RR-intervals  fluctuate in a complex manner. The irregular behavior has motivated researches \cite{Peng,Nature} and still attracts their attention \cite{PengGoldberger,Lin,Struzik,Physica04,Brazilians} to apply modern tools of statistical physics to uncover hidden dependencies in the series, especially those manifested by long-range power-law correlations. 

The well known  indicator of the power-law correlation is the $\beta$ exponent describing decay of the power spectrum $S(f)$.  The $ S(f) \propto (1\slash f)^\beta$ scaling with $\beta =1$  is recognized as manifestation of the fractality of a signal. For many physiological signals, including $RR$-series $1\slash f$-scaling has been found \cite{SpectralScaling, DetStoch}. However, the rigorous meaning of fractality of a signal is related to the Hurst exponent  \cite{FractalMath,Riedi}. The Hurst exponent, called also roughness exponent, traditionally measures the statistical dependence of extravagances in a series on the scale $a$ \cite{Hurst}. But the modern definition of $H$ links the roughness with the notion of self-similarity of the process $X(t)$, i.e.,  with  statistical equivalence of the two difference processes: $\{ X(t + a\delta ) -X(t)\}$  and $ \{ X(t + \delta) -X(t)\} $, for some $a, \delta >0 $. The equivalence means the probabilistic equivalence of the following measures :
\begin{equation}
 X(t + a \delta) -X(t) \equiv_{prob} a^H  (X (t+\delta)-X(t))
 \label{hurst}
\end{equation}
The Hurst exponent $H$ being the distance  dependent value,  expresses  the so-called  long-range dependence   embedded in a time series. 

Moreover, modern statistical physics provides the method, called detrended fluctuation analysis (DFA), \cite{Peng} which also detects long-range correlations.  The power-law relation between the  mean fluctuation of the integrated and detrended subseries and length of the subseries  provides the DFA-exponent $\alpha$. It has been shown that the DFA exponent $\alpha$ differentiates healthy hearts from hearts with  a failure \cite{Peng,Physica04,neuron, Physica1999}. 

Ubiquity of scale invariant properties emerging from  heart rate signals supports the conjecture that RR- series have stochastic rather than deterministic origin since they result from microscopic interactions among many individual components driven by competing forces  \cite{Teich,DetStoch,Nature,Physica04,GoldbergerPeng}. To describe and quantify the mechanisms of these `nonhomeostatic' features, techniques derived from complexity theory: fractal analysis and nonlinear dynamics, are employed. Especially, multifractal analysis provides the resolution of a global exponent to local scaling exponents. It is supposed  that a wide range of temporal and spatial scales enable the organism to adapt to the everyday stressful life \cite{GoldbergerPeng}.

At present what it has been found is that the heart rate of healthy humans is  a multifractal for which  nonzero fractal dimensions and local Hurst exponents are in the interval $h\in(-0.1, 0.5)$ while records for patients with nearly terminal pathology, namely, congestive heart failure, show a significant loss of multifractal complexity by displaying a smaller range of values $h$ \cite{Physica1999,MeyerStiedl}. One says that with the heart dysfunction the transition from multifractal to monofractal scaling in the heart variability takes place. Physiologically, this transition is related to the dysfunction of the control mechanisms regulating the heartbeat --- the autonomous nervous system.

The autonomous nervous system consists of two parts: sympathetic and parasympathetic which act oppositely to the heart rate. The heart rate is  accelerated by the sympathetic system and slowed by the parasympathetic system. Basing on  understanding of competing forces gained by critical phenomena study,  the origin of heart rate complexity is searched in the intrinsic dynamics of this physiological regulatory system. 
Fractal scalings in  heart rate requires the existence of and intricate balance between antagonistic activity of parasympathetic and sympathetic system. The local multifractal scaling in heart rate effects from the interaction between the activity of sympathetic and parasympathetic nervous systems. 

Congestive heart failure is known to be associated with both increased sympathetic system activity and decreased parasympathetic system activity what denotes strong dysfunction of the regulatory system. This  disease reveals as a loss of multifractal scalings in the heart rate series. Moreover, if the parasympathetic blocker - atropine, is served to a subject with the healthy heart then a collapse of the multifractal spectrum is observed while heartbeat dynamics during sympathetic blockade displays a small change toward the multifractal spectrum \cite{PRL}. Furthermore,  RR-series of  patients with clinically recognized autonomic system dysfunction (loss of sympathetic neurons), though display substantially reduced variability to the levels close to congestive heart failure, the conservation of multifractal properties is observed \cite{Struzik}.

The question posed by us is  whether subjects with left ventricular systolic dysfunction (mean LVEF $=30.2\pm 6.7 \%$, NYHA class I--III) is associated with the failure of the autonomic antagonistic activity between parasympathetic and sympathetic nervous systems. 

To achieve the goal, we have carefully choose the group of individuals with  left ventricular systolic dysfunction and the control group of people with the similar age characteristic and without known any cardiologic history. To both groups the two methods of the multifractal analysis is applied: wavelet transform modulus maxima (WTMM) \cite{BacryMuzyArnedo} and multifractal detrended fluctuation analysis (MDFA) \cite{MDFA}. Moreover, we perform the same multifractal analysis to the well-known and widely investigated series  from PhysioNet \cite{physionet} of healthy people, so-called  normal sinus rhythm signals, and  with near terminal pathology, called congestive heart failure signals. So that  we obtain the opportunity to get an insight into the changes of regulatory nervous system caused not only by  heart disease but also by advancing in years.

The contents of the paper is organised as follows. In Section 2 the multifractal formalism is introduced. Section 3 contains the description of numerical methods used by us as well as tests for them are presented. The investigations on heart rate series are in Section 4. The results are shown in a series of figures and carefully discussed. The last section, Section 5, concludes our considerations.

\section{Multifractal analysis}
Multifractal formalism is the way to study scale invariance in their whole variety and dependences \cite{Mandelbrot,FractalMath,Riedi}. The easily recognizable sign of the scale invariance of a  $X(t)$ process  is the occurrence of scaling in any of its statistics, called {\bf a structure function}.

For example, let us consider as the structure function the  statistics of $q$-moments of increments of $X(t)$ in a scale $a$, namely, $ <|X(t+a)-X(t)|^q > $. If :
\begin{equation}
 <|X(t+a)-X(t)|^q >\quad \propto \quad |a|^{\tau(q)}
 \label{partition}
\end{equation}
then an analysis of scaling phenomena means estimates of the corresponding scaling exponents $\tau(q)$ (called {\bf a partition function}). A stochastic processes exhibiting scaling in their statistics is called a fractal stochastic process. 

Depending on relation between $\tau$ and $q$, a fractal process is called:
\begin{itemize}
\item {\bf monofractal }--- if a partition function $\tau(q)$ is linear: $qH-1$. Notice, that the coefficient in this dependence is the Hurst exponent.
\item {\bf multifractal }--- any other dependence between the partition function $\tau$ and the moment $q$.
\end{itemize}

On the other hand, the differentiability of any continuous stochastic process  can be considered. Roughly speaking, a process $X(t)$ is differentiable at point $t_0 $ if there is  a polynomial $P_{t_0}$, usually the Taylor polynomial of $X(t)$ at $t_0$ is considered, for which the following relation holds, \cite{Riedi}: 
\begin{equation}
|X(t)-P_{t_0}(t)| \propto |t-t_0|^{h(t_0)} \quad {\rm for} \quad t\rightarrow t_0
\end{equation}
This relation  gives rise to the notion of the local H\"older exponent $h(t_0)$ describing the degree of  differentiability of a signal at time $t_0$.  It appears that a typical feature of  self-similar processes is that they have a non-integer degree of differentiability, often changing from point to point. Therefore, it is often said that fractal processes are characterized by the inherent, ubiquitous occurrences of irregularities. 

Having estimated $h(t)$ for all $t$, the analysis yields a multifractal spectrum $D(h)$ that may be viewed as a measure of spikiness of the singularity structure of the data. More precisely, if 
\begin{equation}
S(h)= \{ t: h(t) =h \}
\end{equation}
is the subset of time interval where the local singularity exponent is equal to $h$, then the Hausdorff dimension of this set :
\begin{equation}
D(h)= \dim_H S(h)
\end{equation}
assigns the spectrum value $D(h)$ to the $h$ singularity. The partition function $\tau(q)$ and the singularity spectrum $D(h)$ are mutually related functions by the Legendre transform:
\begin{eqnarray}
D(h) &=&\min_q (qh-\tau(q))= q\tau'(q) - \tau(q) \\
\tau(q)&=&\min_h(qh-D(h))= hD'(h) -D(h)
\label{Legendre}
\end{eqnarray}

The most prominent example of a fractal process is the fractional Brownian motion $B_H (t)$ \cite{fBm}.  Since at any scale $a$ the following self-similarity relation  holds, compare (\ref{hurst}) and (\ref{partition}):
\begin{equation}
B_H(a t) \equiv_{prob} a^H B_H(t)
\end{equation}
then  all local H\"older $h(t)$ exponents are independent of  $t$  and $h(t)=H$ for every $t$. Hence $H$, the Hurst exponent,  determines differentiability of the Brownian process.  

There exists a mathematically rigorous approach to  fractional Brownian motions \cite{FractalMath,fBm,Riedi} and  to their generalizations, to the, so-called random cascades and process of the compound Poisson cascade \cite{CPP} which explains fractality by multiplicative organization of a signal. Multiplicative cascades are built from iterative split/multiply procedures that hence produce interdependencies between the different scales of the resulting process.

In case of  real signals arising from fully developed turbulence \cite{turbulence,ParisiFrish85}, financial time series \cite{Mandelbrot,finance} or load of network traffic \cite{traffic},  such local property such as differentiability  can be directly related. The H\"older singularity spectrum is accessible by, for example, the wavelet analysis \cite{BacryMuzyArnedo}. 

However,  a sequence of events extracted from human heart electrocardiogram is different from signals originally studied by multifractal formalism. The notion of differentiability does not make sense here since the fine scales are not accessible. Therefore by  the multifractal formalism  one searches for the spectrum of rough  singularities, so-called, local Hurst exponents \cite{Physica1999,Nature,Physica04}.

\section{Numerical procedures and tests}

Numerical computations of the singularity spectrum for empirical data, straight from the definition, is obviously not feasible. The local singularity exponents vary widely from point to point making their numerical measurement extremely unstable. The way out consists in obtaining the desired multifractal spectrum via  carefully designed the structure function. The following two basic structure functions  are developed to study nonstationary signals and to determine the partition function:
\begin{itemize}
\item[WTMM]--- Wavelet Transform Modulus Maxima\cite{BacryMuzyArnedo}. \\
Here the multiresolution analysis of a signal is done by the wavelet transform.
For a wavelet $\psi(t)$ centered at time zero and scaled $a$, the wavelet coefficient:
\begin{equation}
W(a, t_0)= \frac{1}{a}\sum_{t=1}^N X(t)\psi( \frac{t-t_0}{a})
\end{equation}
measures the signal content around time $(t-t_0)\slash a$. The wavelet is chosen orthogonal to the possible trend. If the trend can be represented by polynomial, a good choice for $\psi(t)$ is the $m$th derivative of a Gaussian $\psi^{(m)}= d^m \slash dx^m \quad e^{-\frac{x^2}{2} } $. The transform eliminates trends up to ($m-1$)-th order. This way the wavelet decomposition contains considerable information on the fractional singularities. 

Then as the structure function the average of local maxima of $|W(a,t)|$ as a function of $a$ is chosen:
\begin{equation}
Z(q, a) = \sum_{local\atop maxima} |W(a,t)|^q \propto a^{\tau(q)}
\label{wtmm}
\end{equation}
 to extract and amplify extravagances in a series.

\item[MDFA]--- Mulifractal Detrended Fluctuation Analysis \cite{MDFA}\\
Similarly to DFA method the local trend $Y^{(j)}$ for each $j$-th subinterval of $a$-th length  is determined and the fluctuation for the subinterval is calculated separately:
\begin{equation}
F^2_j(a)= \frac{1}{a}\sum_{k=(a-1)j}^{aj} [ Y(k) -Y^{(j)} (k)]^2
\end{equation}
Then the structure function means the average over all subintervals of the $q$th order fluctuation function:
\begin{equation}
F(q,a)= \{ \frac{1}{(N/a)}\sum_{j}[F^2_j(a)]^{q/2 } \}^\frac{1}{q}  \propto a^{1+\frac{\tau(q)}{q}}
\label{mdfa}
\end{equation}
where $\tau(q)$ has the same meaning as in (\ref{wtmm}).
\end{itemize}

For certain values of $q$ the exponents $\tau(q)$ have familiar meaning. In particular, $\tau(2)$ is related to the scaling exponent $\beta$ of the power spectrum as $\beta=2+\tau(2)$ and to the Hurst exponent 
$H={1 \slash 2}(1+\tau(2))$. For positive $q$, the structure function describes scaling properties of large fluctuations and strong singularities. For negative $q$ the scaling of small fluctuation and weak singularities is shown. Therefore the partition function $\tau(q)$ by crossing $q=0$ reveals different aspects of cardiac dynamics \cite{Physica1999}.

Since in our calculations we use the software accessible from PhysioNet \cite{physionet}, namely, we work with two packets: DFA.C (prepared by J.Mietus, C-K Peng, and G. Moody ) and MULITIFRACTAL.C (prepared by Y. Ashkenazy) we have to perform tests for the methods abilities. The data sets consisting of 20 000 points with known fractal properties are analyzed in two ways: WTMM (with the 3rd derivative of a Gaussian) and MDFA, both in the q-interval $[-10, 10]$ with the step $\Delta q=0.1$.

\begin{figure}
\begin{center}
\includegraphics[width=0.85\textwidth]{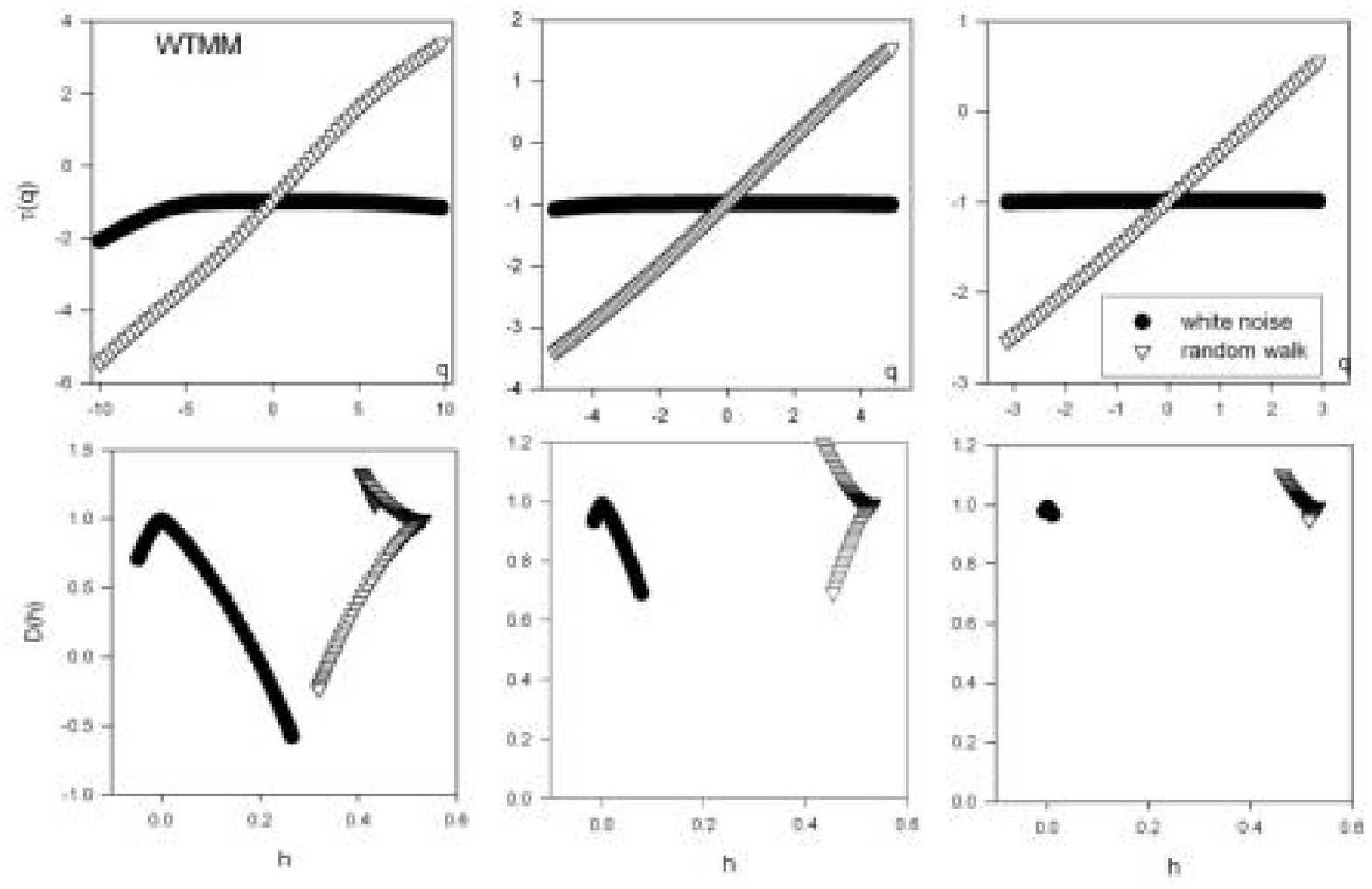}
\includegraphics[width=0.85\textwidth]{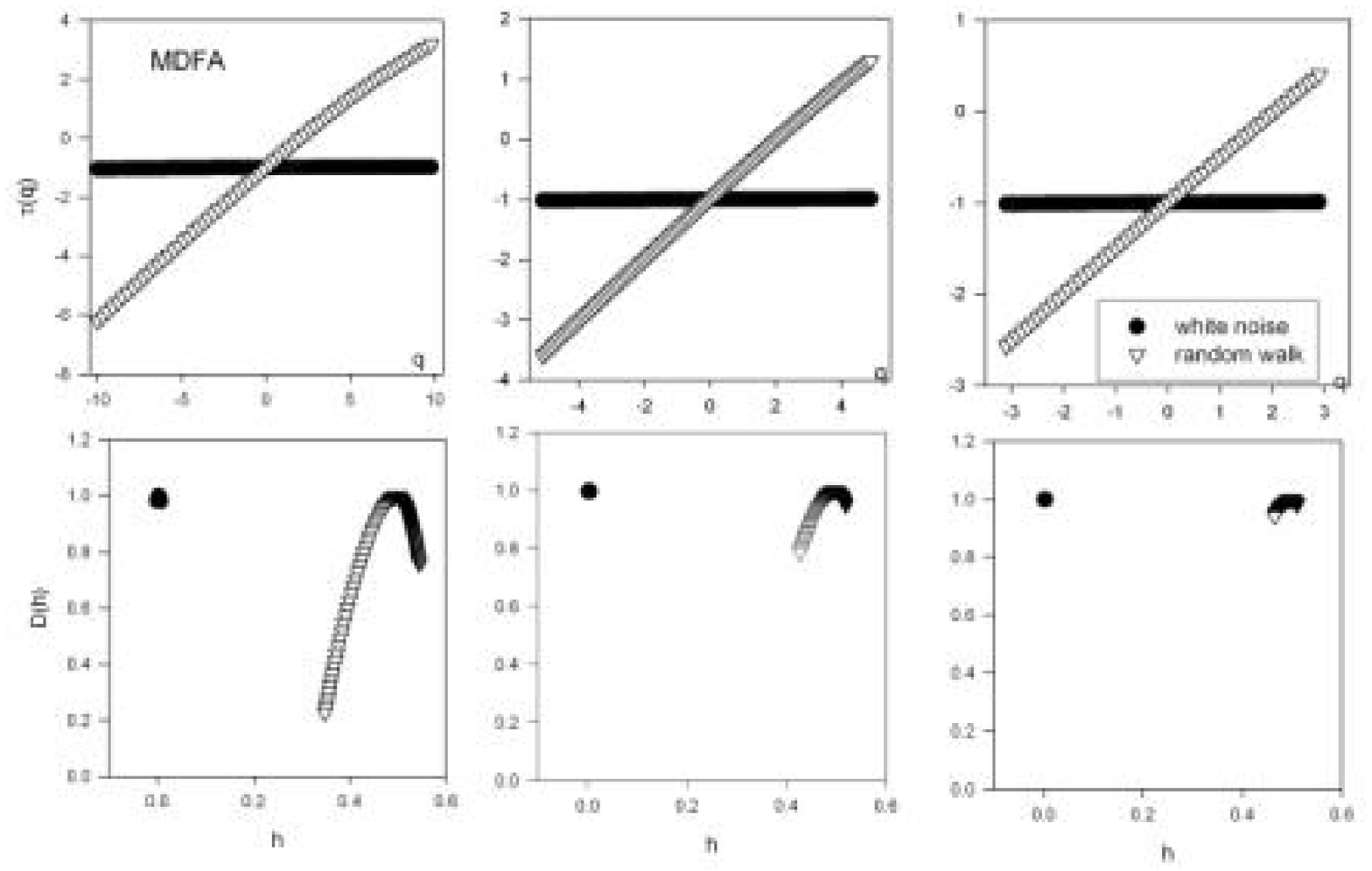}
\end{center}
\caption{\label{pure}Dependence on $q$ of multifractal spectra in case of a white noise and a random walk for WTMM (top figure) and MDFA (bottom figure). In the subsequent figures the q-interval is decreased.}
\end{figure}

The following tests  have been performed: 
\begin{itemize}
\item Multifractality of white noise and random walk, see Fig. \ref{pure}.
\end{itemize}
The series are generated by lrand48() function accessible in ANSI C on UNIX workstations.
The expected result is the linear partition function $\tau(q)$ with the coefficients $0$ for a white noise and  $0.5$ for a random walk. One can notice a large departure  from the theoretical  predictions if the partition functions are considered in the large q-interval. Especially, the wrong results are obtained when $q$ is negative and WTMM method is applied.  The MDFA method provides  correct values in the whole q-interval  in case of a  white noise, namely, the spectrum is a point : $\{ (0,1)\}$. But in case of a random walk, the point-like spectrum  is satisfactory only when the q-interval is shrunk , e.g.,  to  $|q|<3$.

\begin{figure}
\begin{center}
\includegraphics[width=0.9\textwidth]{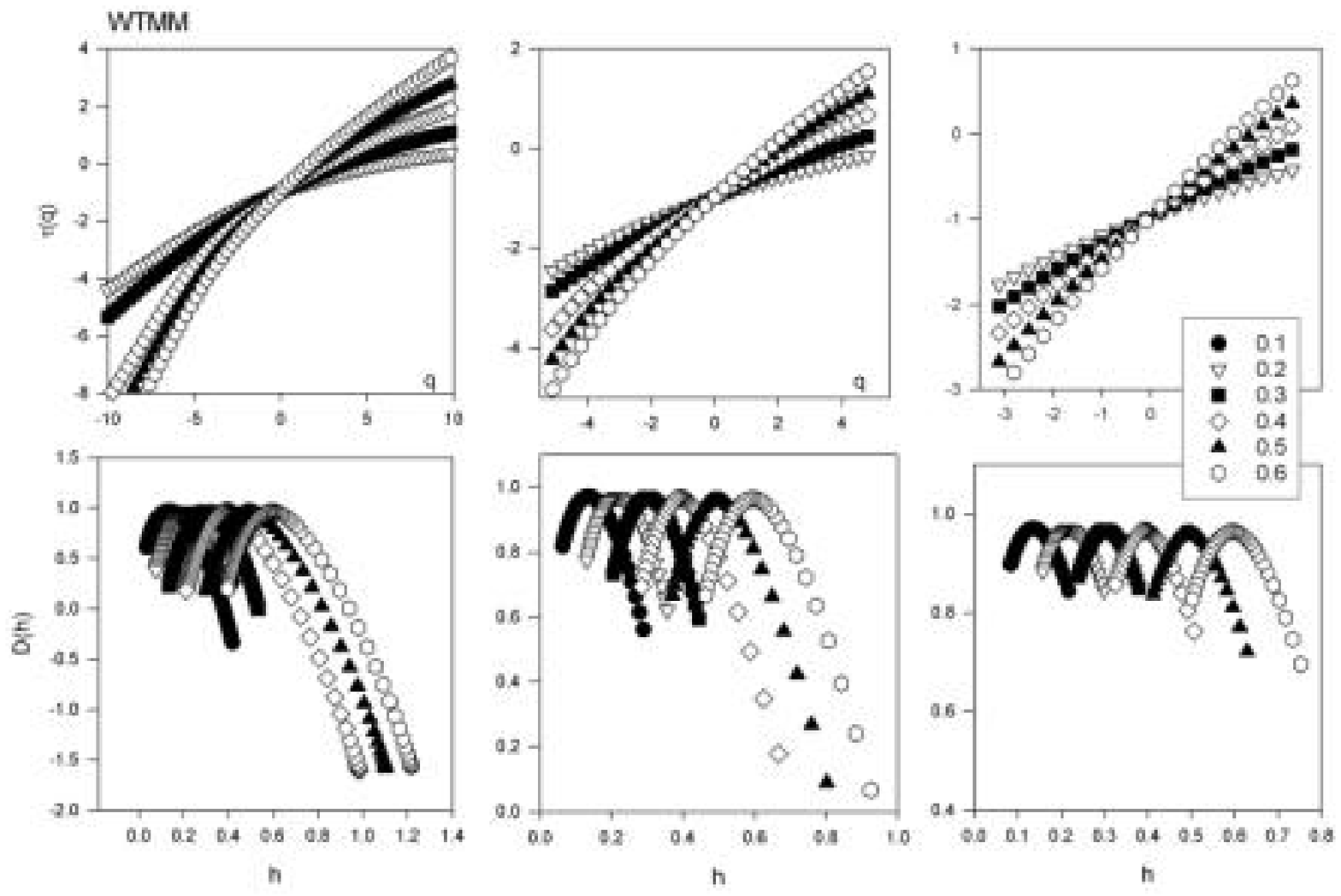}
\includegraphics[width=0.9\textwidth]{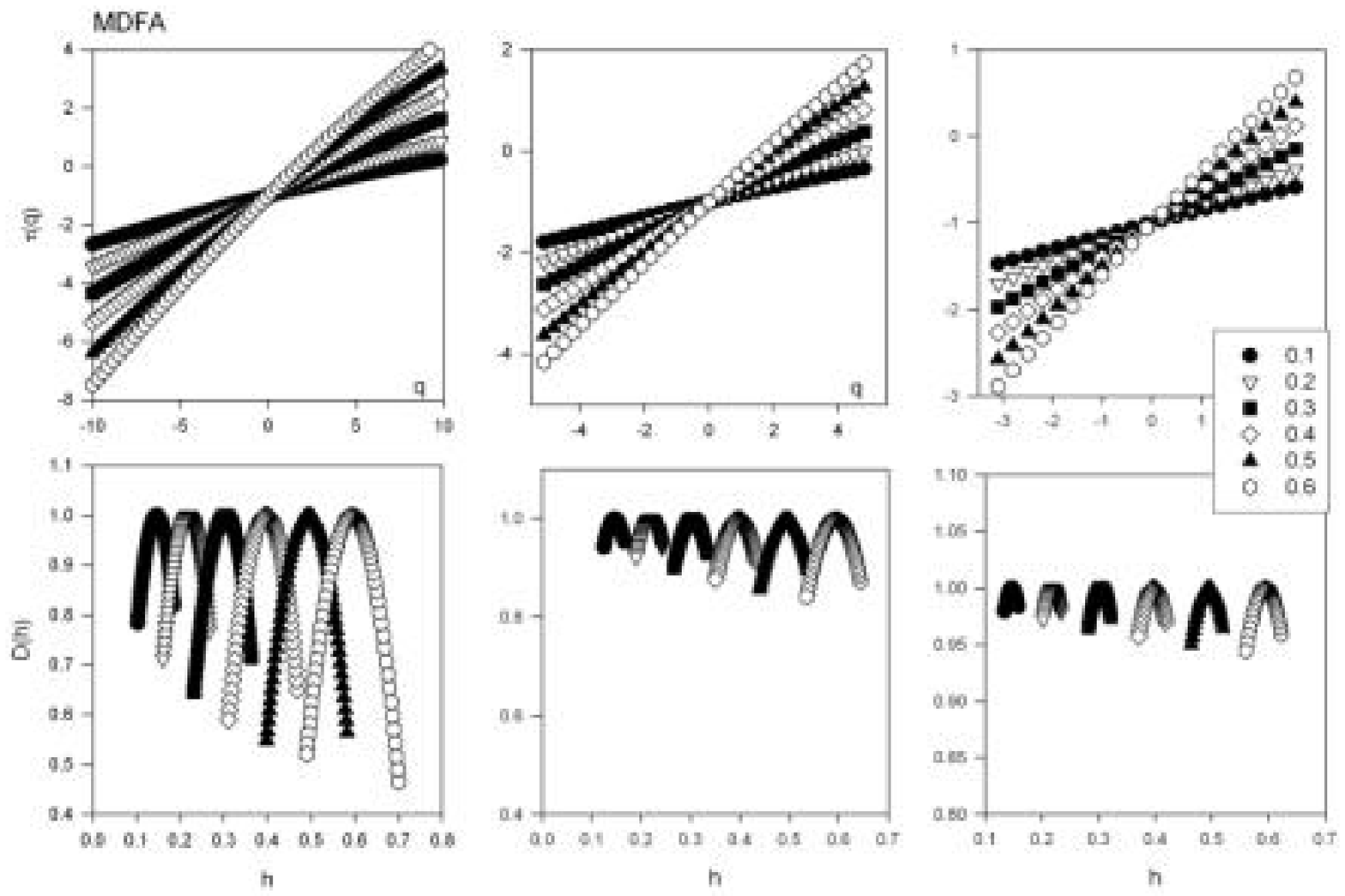}
\end{center}
\caption{\label{frac}Dependence on $q$ of multifractal spectra in case of fractal Brownian motions if WTMM (top figure) and MDFA (bottom figure) methods are applied. In the subsequent figures the q-interval is decreased. Different point markers correspond to different Hurst exponents as denoted by the labels.}
\end{figure}

\begin{itemize}
\item Multifractality of a fractional Brownian motion, see Fig. \ref{frac}. 
\end{itemize}

The series are  generated with the help of  {\it tsfBm} packet \cite{tsfBm}. From Fig. \ref{frac} one can  read that the partition functions are not linearly depended on $q$, especially, when WTMM method with  negative $q$ is calculated. Therefore the corresponding  spectra are not  point-like,  they  are not symmetrical --- they have wide right wings. Sizes of these spectra are related to the method.  MDFA method leads to results which systematically diverge from the expected linear behavior as $|q|$ grows. When restricting the q-interval then both methods provides  the expected shape, namely,  a point spectrum $\{(H,1)\}$. 

\begin{figure}
\begin{center}
\includegraphics[width=0.9\textwidth]{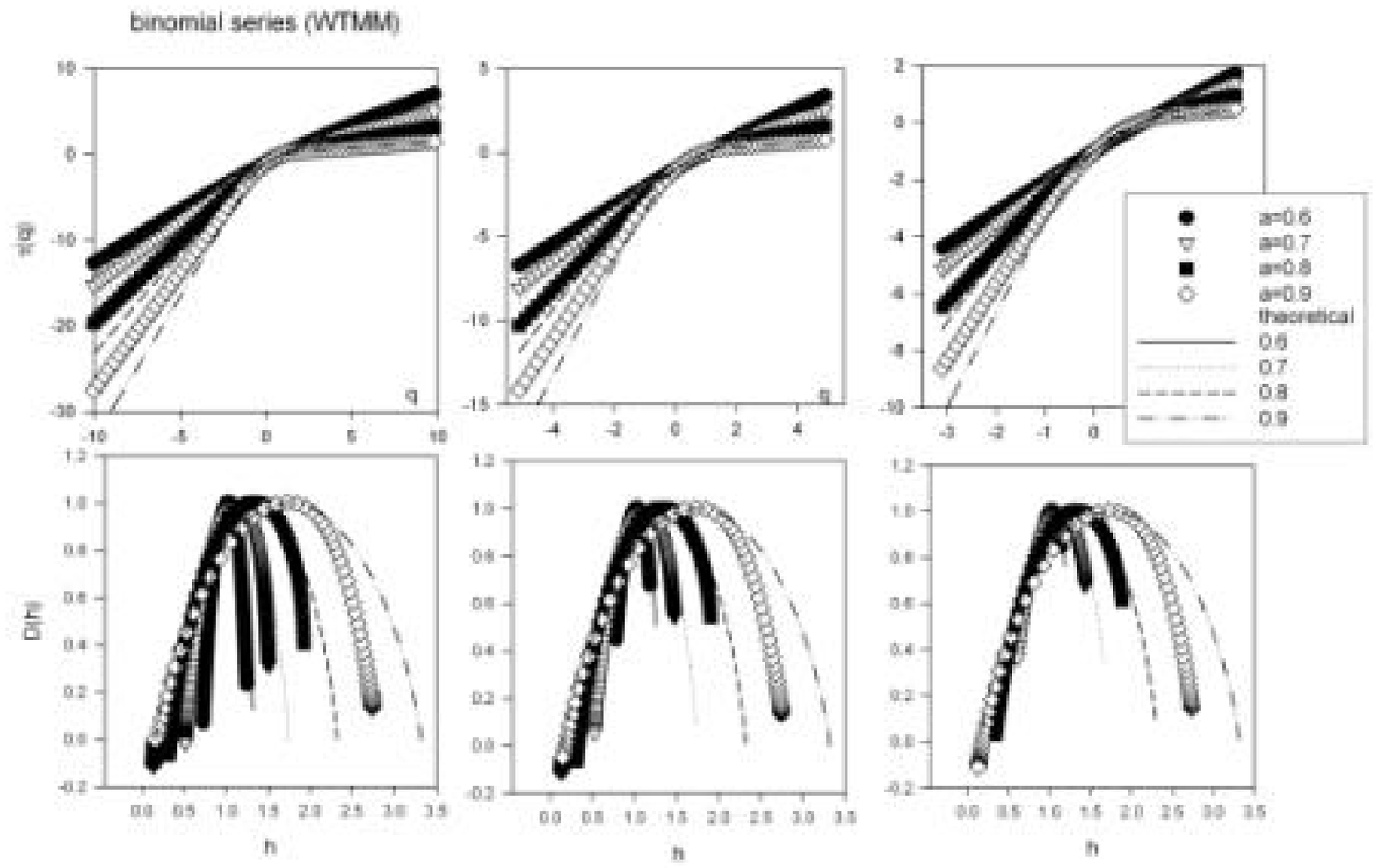}
\includegraphics[width=0.9\textwidth]{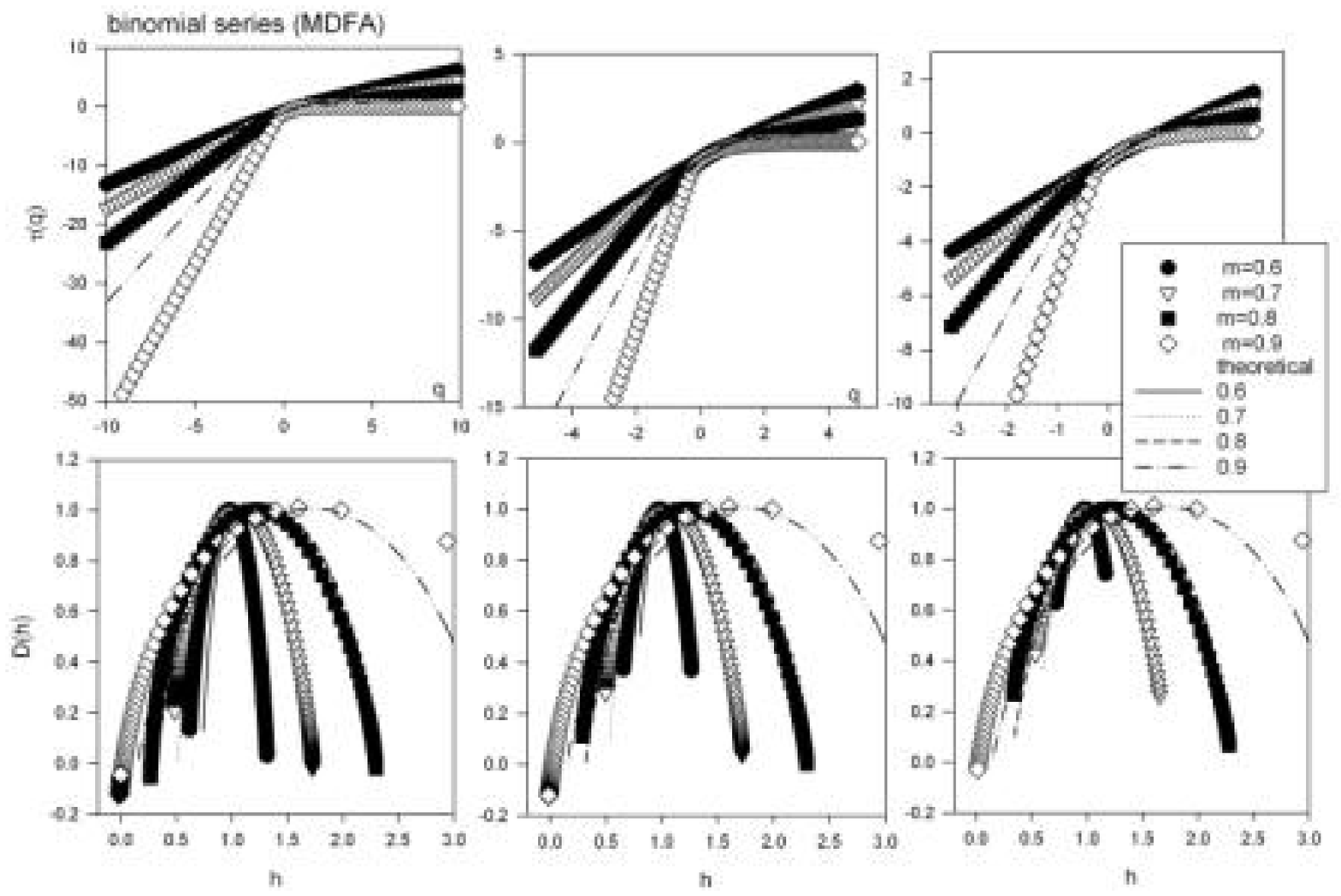}
\end{center}
\caption{\label{binomial}Dependence on $q$ of multifractal spectra in case of binomial measures with different mass values and  for WTMM (top figure) and MDFA (bottom figure). In the subsequent figures the  q-interval is decreased. }
\end{figure}

\begin{itemize}
\item Multifractality of binomial series, Fig. \ref{binomial}, \cite{Mandelbrot}
\end{itemize}
The binomial measure is the simple example where a strictly multifractal spectrum appears. The binomial series represent a binomial cascade measure. These cascades are constructed through the following multiplicative iteration schema: Let divide the interval into two parts and assign a mass $m$ to the right part and $1-m$ to the left part of the division. Then, repeat in each iteration, each of the interval divide into two parts and assign for the right part: the whole interval mass multiplied by $m$, and for the left part: the whole interval mass multiplied by $1-m$.

The theoretical approach gives the following values for $\tau(q)$
\begin{equation}
\tau(q)= \frac{-\ln [ m^q + (1-m)^q]}{\ln 2}
\end{equation}

Both methods  WTMM and MDFA give unreliable results if $q < 0$ --- WTMM method underestimates the theoretically calculated  $\tau$ values while MDFA method overestimates these theoretical values what effects in shrinking (WTMM) or enlarging (MDFA) the spectra. On the other hand, for positive  $q $, both methods reveal the expected properties.  Noticeable, that independently of the q-interval the large convex spectrum shape is preserved.

\section{Heart rate study}
The multifractal analysis of RR-series is  done for the data downloaded from  PhysioNet page \cite{physionet} and  for  RR-series collected and selected by us  and accessible on request \cite{amgdata}. The 24-hours ECG Holter signals are carefully analyzed and annotated to extract normal RR signals.

\subsection{The series}

We analyze four groups of subjects:

--- {\bf nk} group. The {\bf nk} group (the basic group of the study) consists of 90 patients hospitalized during 2001--2004 years in the 1st Department of Cardiology of Medical University in Gda\'nsk, Poland (9 women, 81 men, the average age is $57 \pm 10$) in whom the reduced left ventricular systolic function was recognized by echocardiogram  due to the low left ventricular ejection fraction (LVEF $\le 0.40 \%$, mean LVEF $=30.2\pm 6.7 \%$). The additional criteria which exclude subjects from the {\bf nk} group are: the myocardial infarction in last 6 months, persistent atria fibrillation, sinus-node disease, recognized diabetes mellitus, coronary revascularisation in last 6 months or kidney failure with creatine level $ >2$.

--- {\bf gk} group. The control group is made of 40 healthy individuals (4 women, 36 men, the average age is $52 \pm 8$) without past history of cardiovascular disease, with both echocardiogram and electrocardiogram in normal range. The left ventricle ejection fraction was normal (mean LVEF $= 68.0 \pm 4.7 \%$).

For each person from the above two groups the 24-hour ECG Holter monitoring was performed. The signal was digitized using Delmar Avionics recorder (Digitorder) and then analysed and  annotated using Delmar Accuplus 363 system (fully interactive method) by an experienced physician. The minimum number of qualified sinus beats required for the signal  to enter the study is $90\%$.

--- {\bf chf} group. The group consists of 13 patients with congestive heart failure (3 women, 10 men, the average age is $56 \pm 12 $) of whom the whole day data are available from the PhysioNet \cite{physionet} (www.physionet.org$\slash$physiobank$\slash$databas$\slash$chfdb). This severe heart failure is known to be related to the strong dysfunction of the autonomic nervous system.

--- {\bf nsr} group. The group of 14 healthy and rather young people (11 women, 3 men, the average age is $35\pm 8$)  whose ECG recordings are available from the PhysioNet \cite{physionet} (www.physionet.org$\slash$physiobank$\slash$databas$\slash$nsrdb).

The PhysioNet database consists of 18 {\bf nsr} records and 15 {\bf chf} signals. From 24-hour series downloaded from the PhysioNet the normal beats were determined following annotations. Only series with the number of normal annotations greater than $80\%$ are included into considerations.

From each signal the two  5-hour continuous subsets were extracted:  diurnal and nocturnal. These subsets were extracted manually, following the easy recognizable in RR-series the two stages:  daily activity and sleep state. In most cases these periods are consistent with hours 15:00--20:00 for the daily activity and 0:00--5:00 for the sleep state. 

\begin{figure}
\begin{center}
\includegraphics[width=0.9\textwidth]{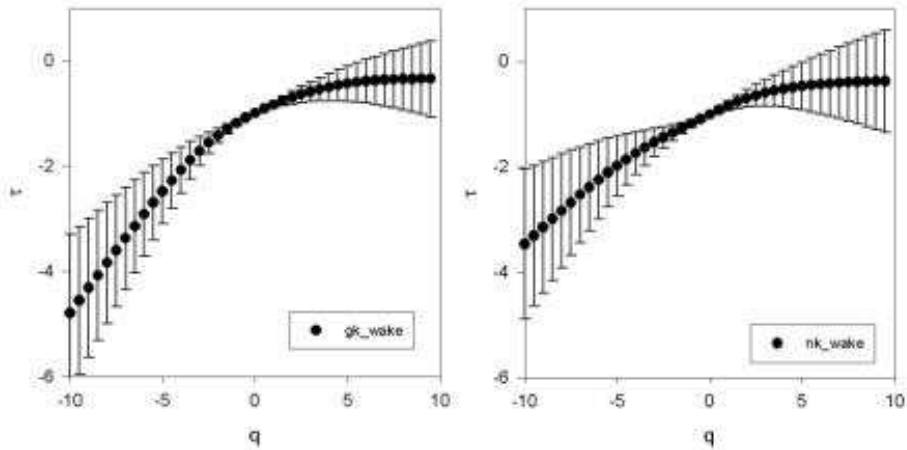}
\end{center}
\caption{\label{errors}Average partition functions together with standard deviation errors obtained when WTMM method is applied to gk\_wake and nk\_wake RR-series.}
\end{figure}

Finally, we deal with 8 groups of series named  subsequently:  {\bf nk\_wake, gk\_wake, chf\_wake, nsr\_wake}, {\bf nk\_sleep, gk\_sleep, chf\_sleep, nsr\_sleep.}
For each series the partition function $\tau(q)$ is calculated by both methods: WTMM  and MDFA in  the interval  $q=-10,\dots, 10$ with a $\Delta q=0.1$ step. The exponents $\tau(q)$ are obtained from  slopes of $Z(q,a)$ by (\ref{wtmm}) in case of WTMM method  and $F(q,a)$ by (\ref{mdfa}) in case of MDFA method for scales $a > 10$. Then the group averaged partition function is found and the multifractal spectrum is calculated following the Legendre transform (\ref{Legendre}). The standard deviation errors of the estimated  partition functions in case of MDFA do not exceed $5\%$ for all $q$ while in case of WTMM method  the errors grow significantly with $|q|$ departuring from 0 as it is shown in Fig. \ref{errors} in case of {\bf gk\_wake} and {\bf nk\_wake} RR-series. Similar results are obtained for other series.

\subsection{The results}
\begin{figure}
\begin{center}
\includegraphics[width=0.9\textwidth]{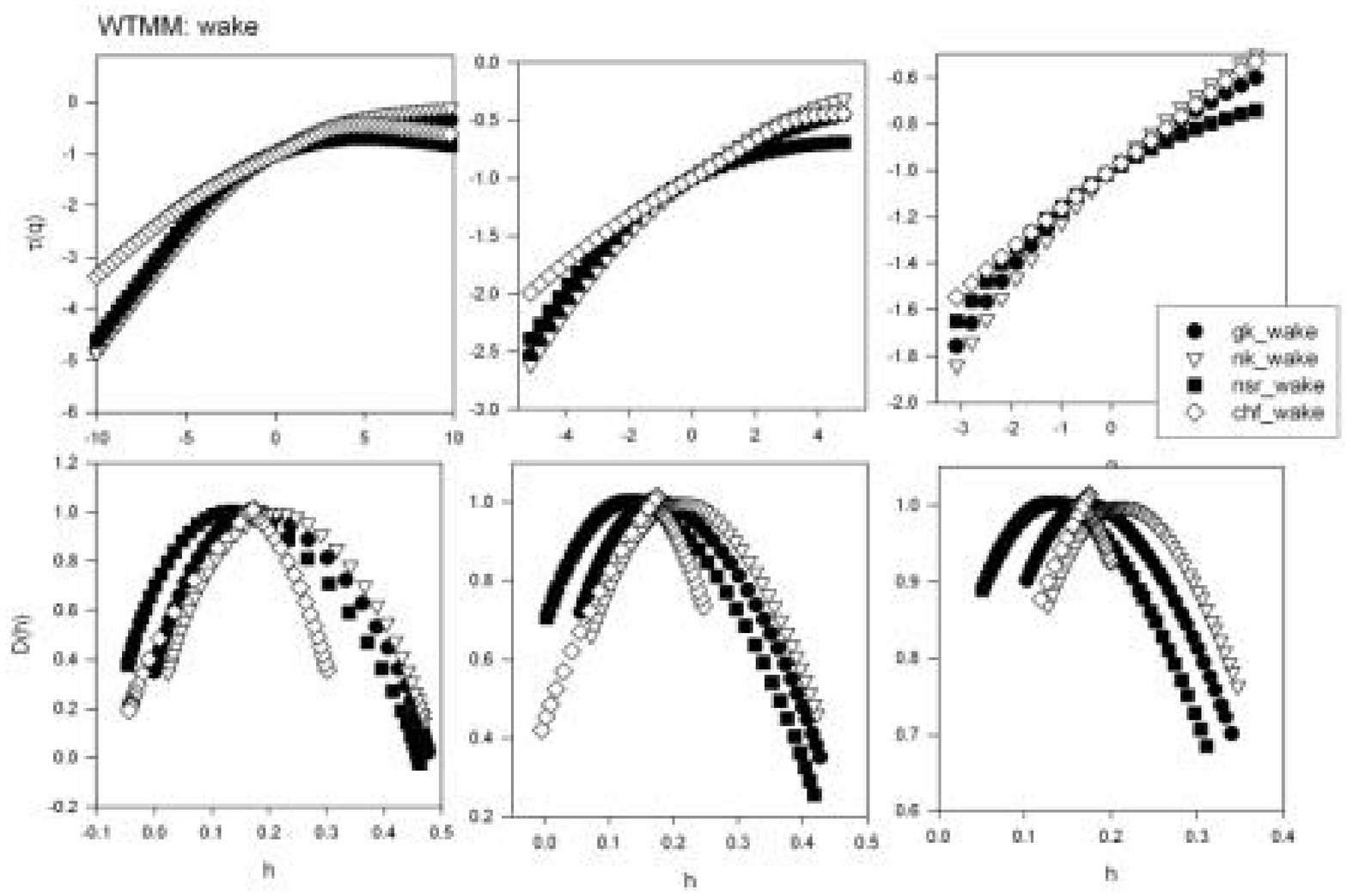}
\includegraphics[width=0.9\textwidth]{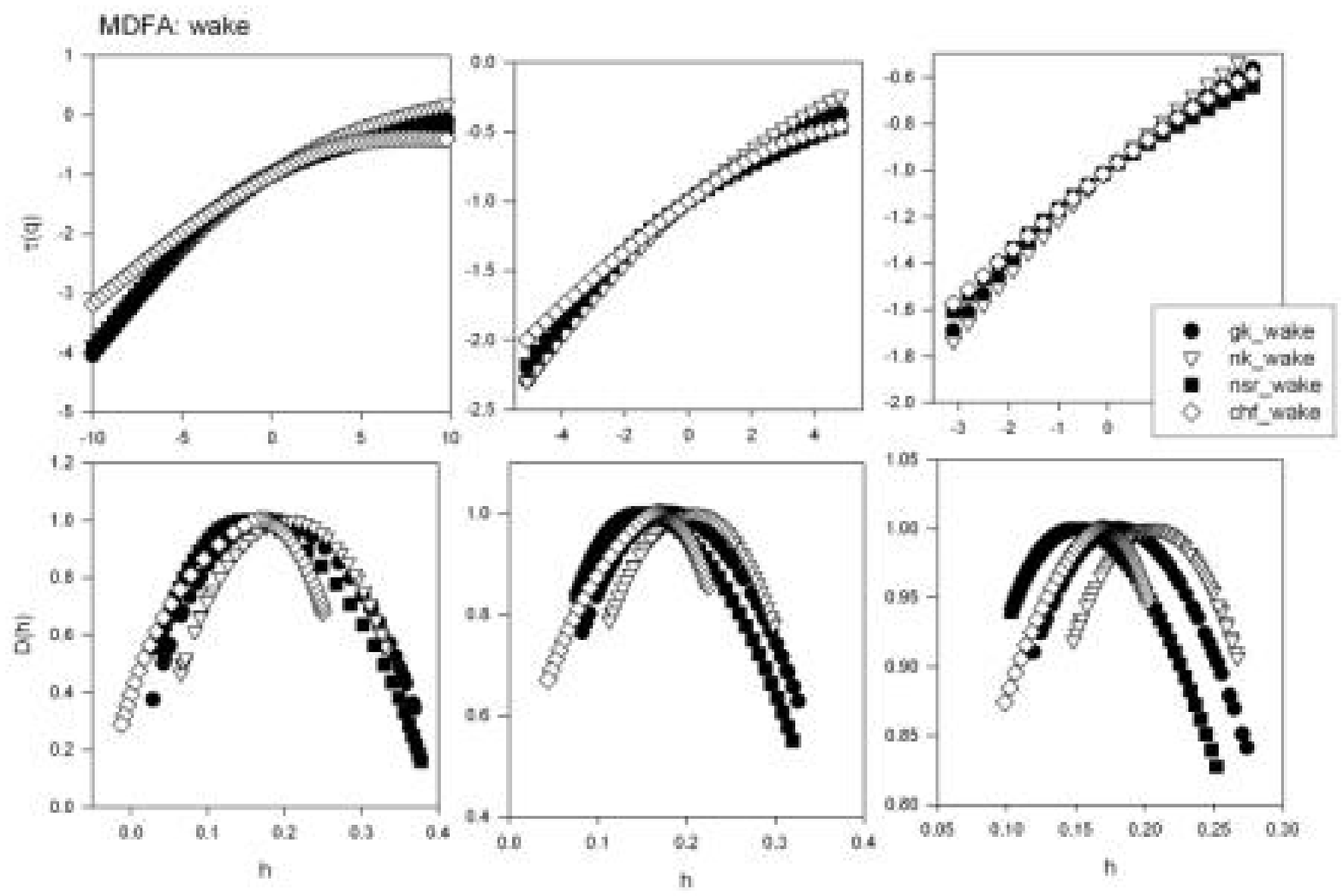}
\end{center}
\caption{\label{wake}Dependence on $q$ of multifractal spectra for wake series  if WTMM (top figure) and MDFA (bottom figure) methods are applied. In the subsequent figures the q-interval is decreased.}
\end{figure}

\begin{figure}
\begin{center}
\includegraphics[width=0.9\textwidth]{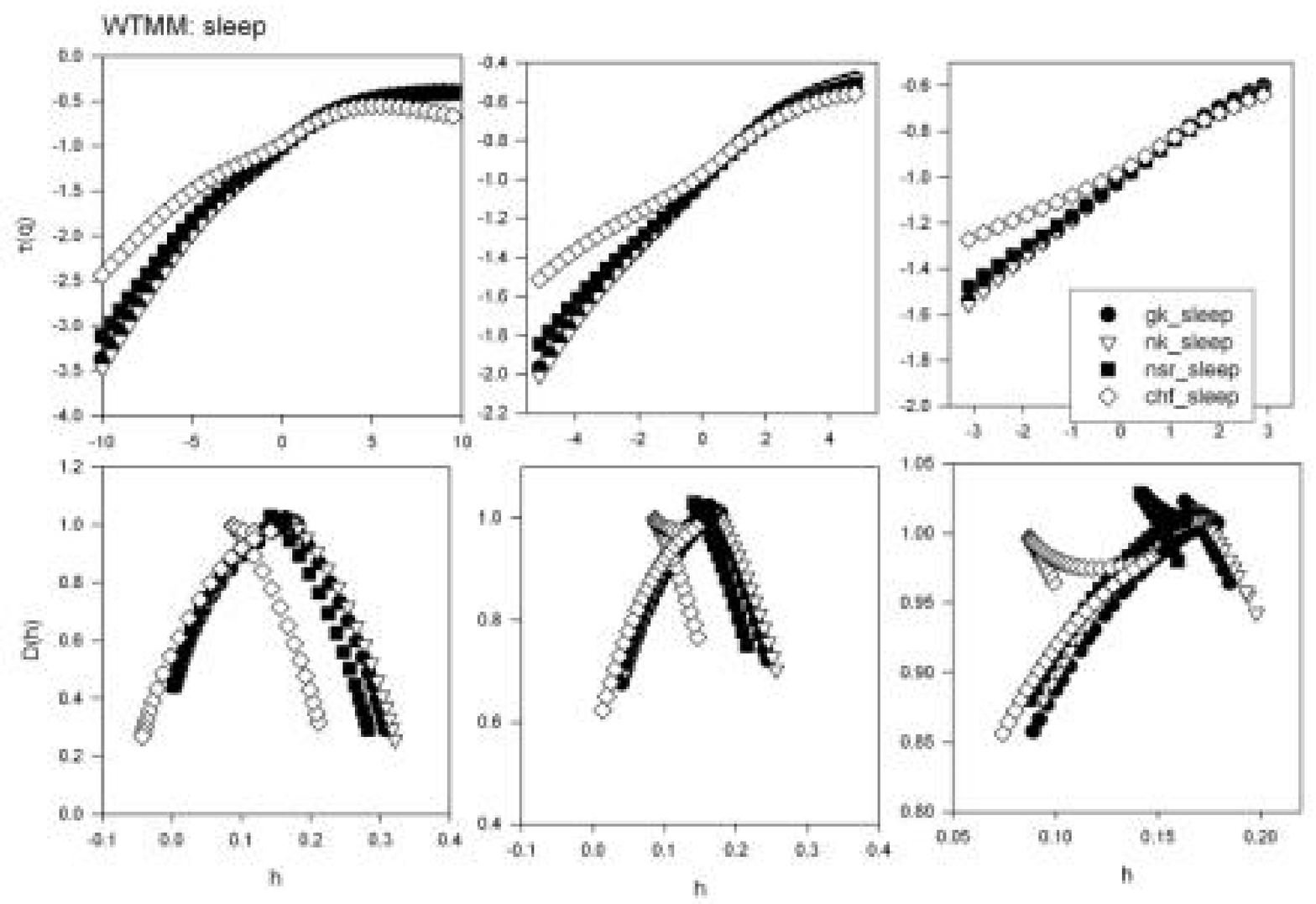}
\includegraphics[width=0.9\textwidth]{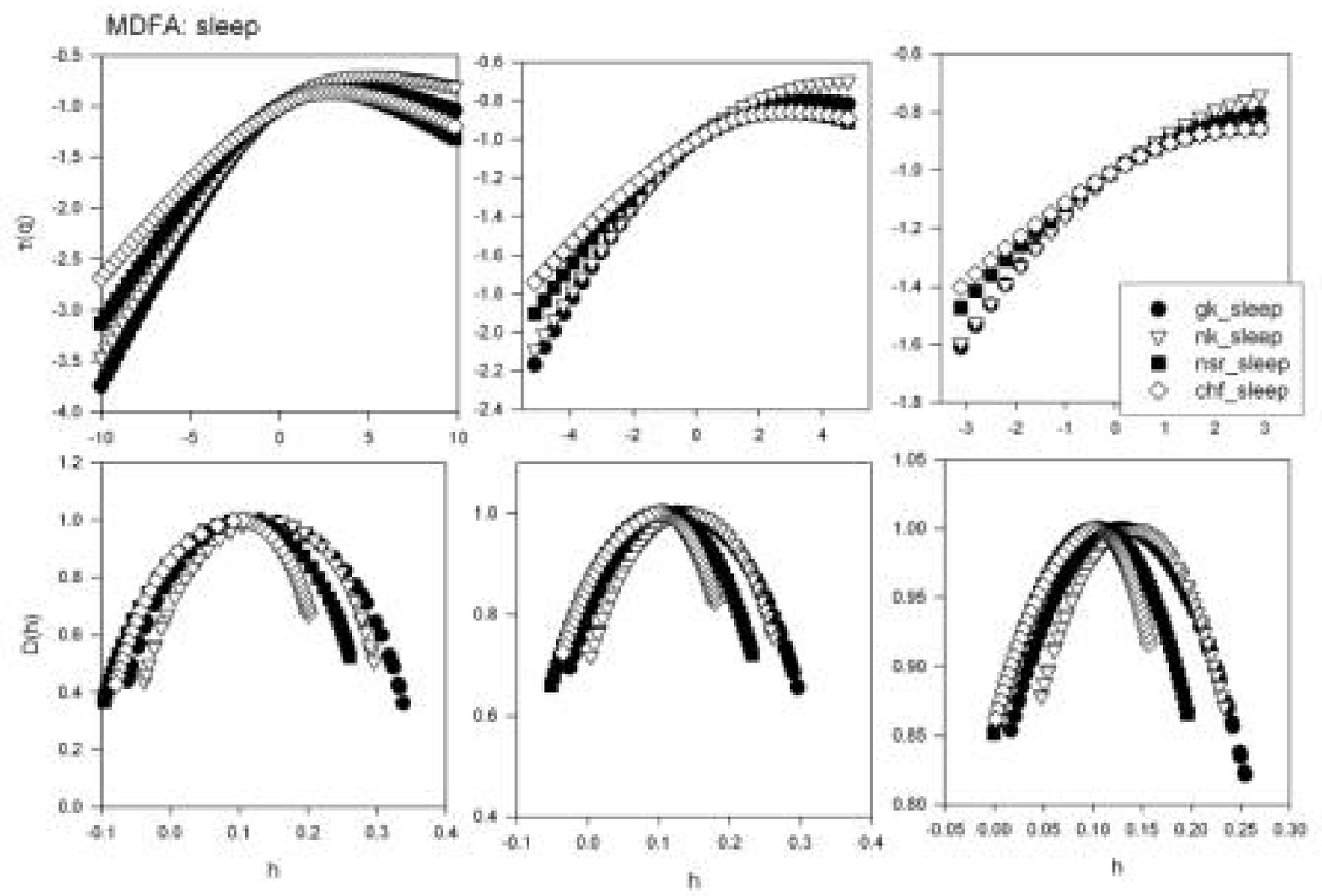}
\end{center}
\caption{\label{sleep}Dependence on $q$ of multifractal spectra for sleep series if WTMM (top figure) and MDFA (bottom figure) are applied. In the subsequent figures the q-interval is decreased.}
\end{figure}

The results of multifractal analysis are presented in Figs. \ref{wake},  \ref{sleep} and in Table 1, 2. We show the results in the whole studied $q$-interval, but our analysis concentrates on $|q|<3$  interval to decrease possible numerical  effects discussed in the previous section.

If  $q$ is  positive then both methods  provide the {\bf nsr\_wake} partition function $\tau_{nsr\_wake}(q)$ distinct from others. The resulting spectrum is the widest.  However,  {\bf gk\_wake} and {\bf nk\_wake} also lead to the wide spectra  but the spectra  locations are moved toward higher $h$ values (compare $h_{l}$, $h_{max}$ and $H$ in Table 1).  The spectrum obtained from {\bf chf\_wake} series is different. It is significantly narrow. The presented results allow us to  claim  that aging and disease influence the heart complexity. The process has two stages. At first, the anticorrelation (feedback) mechanisms are weakened, namely, the response of the autonomic nervous system is reduced, and then, secondly, the autonomic system  response is restricted to few  regulatory mechanisms only.

\begin{table}
\caption{\small Properties of multifractal spectra for  $|q|<3$, daytime series. $h_{l}$ denotes the most left value of the local Hurst exponent noticed in the interval, $D(h_{l})$ means its probability; $h_{r}$ denotes the most right value of the local Hurst exponent noticed in the interval, $D(h_{r})$ means its probability; $h_{max}$ is the value of the singularity exponent when  $q$ crosses the zero. $H$ is the global Hurst exponent calculated as $H={1\over 2}(1+\tau(2))$}
	\begin{center}
		\begin{tabular}{|c|c|c|c|c|c|c|c|c|}
		\hline
		series &$h_{l}$ &$h_{r}$ & total  &$h_{max}$ & $h_{l}$  & $h_{r}$ & total  &$h_{max}$ \\
		 &$D(h_{l})$ & $D(h_{r})$ & width & Hurst & $D(h_{l})$ & $D(h_{r})$ & width & Hurst  \\
		 \hline\hline
		 & \multicolumn{4}{|l|} { WTMM} &\multicolumn{4}{|l|} {MDFA} \\ 
		\hline\hline
		gk\_wake & 0.102 & 0.333 & 0.231 & 0.17 & 0.120 & 0.271 & 0.149 & 0.18  \\
		         & 0.90& 0.72  & & 0.15&  0.91& 0.85  & & 0.16\\     
		\hline
		nk\_wake & 0.125 & 0.341 & 0.216 & 0.21 & 0.146 & 0.267 & 0.121 & 0.21 \\
		         & 0.86 & 0.78 & & 0.19& 0.91 & 0.91 & & 0.19\\     
		\hline
		nsr\_wake & 0.049 & 0.304 & 0.255 & 0.13 & 0.102 & 0.248 & 0.146 & 0.15 \\
		         & 0.88   & 0.71 & & 0.10& 0.94   & 0.84 & & 0.13\\     
		\hline
		chf\_wake & 0.111 & 0.199 & 0.088 & 0.17 &  0.095 & 0.200 & 0.105 & 0.17 \\
		         & 0.85 & 0.93 & &0.17&  0.86 & 0.95 & &0.15\\     
		\hline
	\end{tabular}
	\end{center}
\end{table}

\begin{table}
\begin{center}
\caption{\small Properties of multifractal spectra for  $|q|<3$ in case of nighttime series, MDFA method only. Notation is the same as in Table 1.}
		\begin{tabular}{|c|c|c|c|c|}
		\hline
 		series &$h_{l}$ &$h_{r}$ & total  &$h_{max}$  \\
		 &$D(h_{l})$ & $D(h_{r})$ & width & Hurst  \\
		 \hline\hline
		 & \multicolumn{4}{|l|} {MDFA} \\ 
		 \hline
		gk\_sleep  &  0.013 & 0.254 & 0.241 & 0.13  \\
		           &  0.84& 0.82  & & 0.08\\     
		\hline
		nk\_sleep & 0.041 & 0.230 & 0.189 & 0.14  \\
		          & 0.86 & 0.88 & & 0.10\\     
		\hline
		nsr\_sleep &  -0.004 & 0.194 & 0.198 & 0.11\\
		          & 0.84   & 0.87 & & 0.06\\     
		\hline
		chf\_sleep&   -0.001 & 0.155 & 0.156 & 0.11  \\
		         & 0.85 & 0.92 & &0.07\\     
		\hline
	\end{tabular}
	\end{center}
\end{table}

\begin{figure}
\begin{center}
\includegraphics[width=0.95\textwidth]{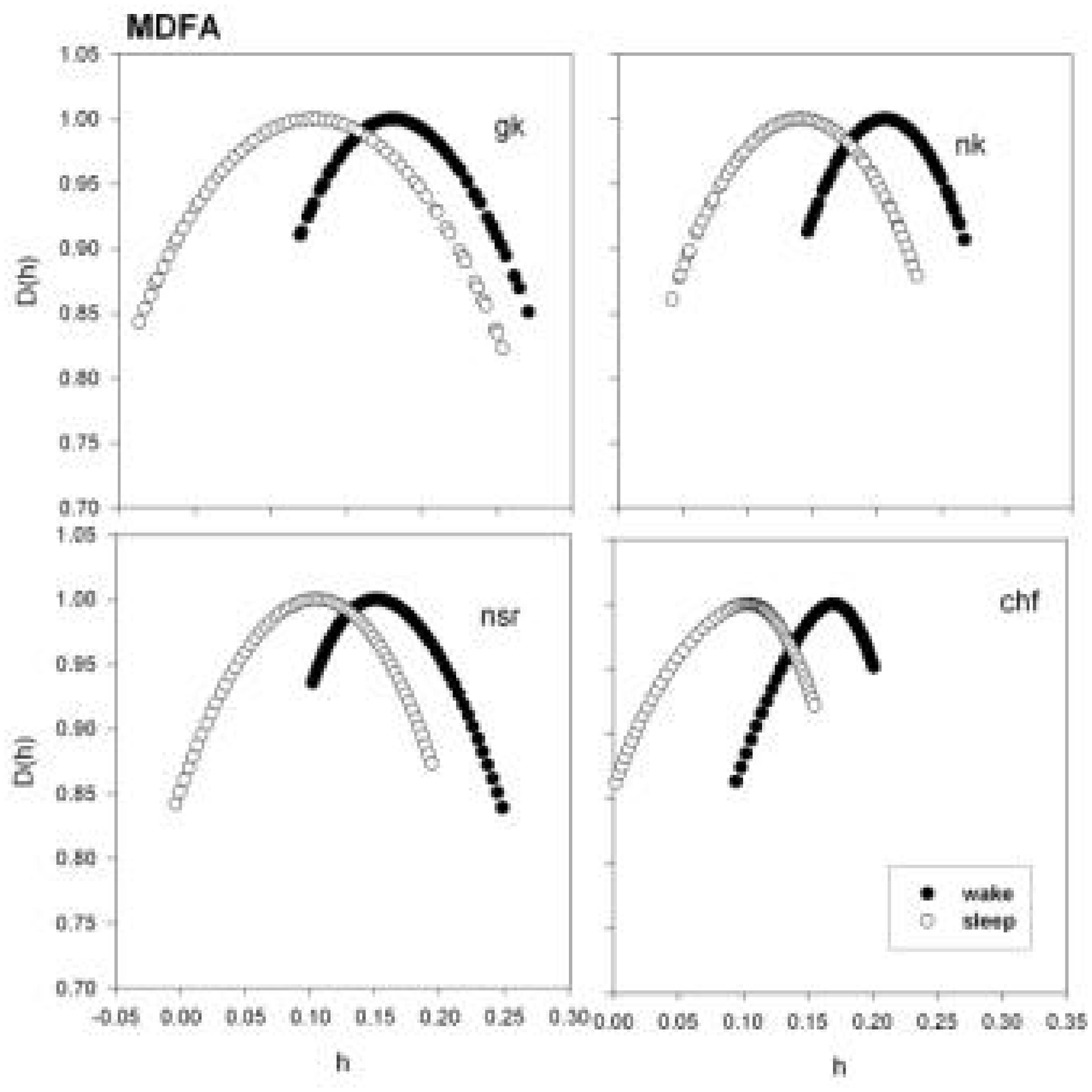}
\end{center}
\caption{\label{mdfa_ws} Multifractal spectra for wake and sleep series if MDFA is applied for $|q|<3 $. }
\end{figure}

Results obtained for the sleeping human state are consistent with the previously observed fact  that multifractality is not reduced at minimised  physical activity such as constant routine protocol, where physical activity and postural changes are kept to a minimum \cite{PRL,Struzik}. The nocturnal activity does not reduce variability of singularity exponents, also. 

The WTMM method leads to  non-convex partition functions  what effects in numerical instability in estimates of the  spectra, though the MDFA method  provides regular shapes, see Fig. \ref{sleep}. Therefore we  present the multifractal characteristics corresponding to the MDFA method only,  see  Table 2.  To amplify difference between diurnal and nocturnal series we show the MDFA spectra in  the subsequent panels of Fig. \ref{mdfa_ws}.
The nighttime spectra are moved to lower $h$ values but  the  spectra are wider than  the spectra of corresponding wake series. The  global Hurst exponent is significantly lower. Hence the  results indicate even increase of the impact of the regulatory system at the sleeping state, and rather support the hypothesis that there are changes in cardiac control during sleep and wake periods which lead to the systematic changes in the scaling properties of the heartbeat dynamics, see e.g. \cite{Physica04,MeyerStiedl}.

\section{Conclusions}
The cardiac time series exhibits wild irregular fluctuations and nonstationary behavior in the healthy subject which is much attenuated in cardiac diseases and/or by aging. 
The question posed by us in this paper  whether the left ventricular dysfunction results in the change of multifractal properties.  

Both methods: WTMM and MDFA show  that the transition in the correlation properties takes place. From strong and  multi-scale anticorrelations --- reach multifractal state, in case of healthy and young hearts, by weakening these properties (less anticorrelation and less multi-scales) caused by  advance in years of a subject, and further loosing control  in case of a subject with left ventricle systolic dysfunction.  Finally, domination of few scales occur  --- a monofractal state, when the heart is seriously damaged with clinical manifestation of the congestive heart failure. 
The spectra of wake series  obtained  by both methods display these properties: the shrink of $h$ interval when we move from the healthy heart to the heart with a disease and the change in the shape of spectrum from parabola-wide shape to triangle-like shape.  
The  nocturnal data provide different property ---  the anticorrelated feature of the heartbeat fluctuations increases during the sleep human state.

We believe that our findings are important in putting forward the role of multifractal analysis for diagnose the conditions a range of patients

{\noindent\bf Acknowledgments}\\
We wish to acknowledge the support of the Ministry of Science and Information Technology KBN PB$\slash$1472$\slash$PO3$\slash$2003$\slash$25

\end{document}